# Predicting nanocrystal morphology governed by interfacial strain – the case for NiO on SrTiO$_3$

Hongwei Liu[1, ‡, *], Xuan Cheng[2, 3, †, ‡], NagarajanValanoor[2, *]

AUTHOR ADDRESS

[1] Australian Centre for Microscopy and Microanalysis, The University of Sydney, Sydney, NSW 2006, Australia

[2] School of Materials Science and Engineering, University of New South Wales, Sydney, NSW 2052, Australia

[3] Department of Materials Science and Engineering, Monash University, Melbourne, VIC 3800, Australia

KEYWORDS: *Nanocrystal morphology; Heteroepitaxy thin film; Interface strain energy; transmission electron microscopy.*

**ABSTRACT:** The shape dependence for the technologically important nickel oxide (NiO) nanocrystals on (001) strontium titanate substrates is investigated under the generalized Wulff-Kaichew (GWK) theorem framework. It is found that the shape of the NiO nanocrystals is primarily governed by the existence (or absence) of interfacial strain. Nanocrystals that have a fully pseudomorphic interface with the substrate (*i.e.* the epitaxial strain is not relaxed) form an embedded smooth ball-crown morphology with {001}, {011}, {111} and high-index {113} exposed facets with a negative Wulff point. On the other hand, when the interfacial strain is relaxed by misfit dislocations, the nanocrystals take on a truncated pyramidal shape, bounded by {111} faces and a {001} flat top, with a positive Wulff point. Our quantitative model is able to predict both experimentally observed shapes and sizes with good accuracy. Given the increasing demand for hetero-epitaxial nanocrystals in various physio-chemical and electro-chemical functional devices, these results lay the important groundwork in exploiting the GWK theorem as a general analytical approach to explain hetero-epitaxial nanocrystal growth on oxide substrates governed by interface strain.

**I Introduction**

Thin film heteroepitaxy has led to the discovery of a vast number of emergent properties in tailor made heterostructures fabricated on oxide substrates.[1] These epitaxial heterostructures are fabricated by the overgrowth of a crystalline layer on a crystalline substrate. In the special case when heteroepitaxy yields individual nanocrystal islands as opposed to continuous thin films, the nanocrystals take a certain shape, which is the consequence of several interfacial and surface-driven phenomena. This interface-controlled morphology is a ubiquitous phenomenon irrespective of the type (vapor/liquid/solid or chemical/physical) of phase deposition.[2] It is primarily governed by the preferred crystallographic orientation relationship and interface structure between the overlayer and the substrate.[3] Furthermore, more complex factors such as gas absorption or chemisorption at the exposed facets or surface reconstruction during growth also play key roles.[4] As such, the ultimate functional performance is shown to be strongly dependent on such shape and surface termination driven changes to the nanocrystals.[5] Therefore the interpretation of why the nanocrystal adopts a certain shape and more importantly the prediction of morphology has thus evolved to been one of the most intensely studied topics of in the design and development of free-standing epitaxial nanocrystals.

Here we demonstrate the ability of generalised Wulf-Kaishew (GWK) theorem by predicting the morphology of the epitaxial nickel oxide (NiO) nanocrystals grown on strontium titanate (SrTiO$_3$, STO) (001) substate We choose this system as it has demonstrated significant technological promise, with applications in photocatalysis,[6] electrochromism,[7] resistive switching,[8] and magnetotransport.[9] We show how the shape and termination of the NiO nanocrystals surfaces is affected by the surface energy, the interfacial strain energy raised from lattice mismatch, as well as energy released due to the presence of dislocations. These energy factors are explicitly computed and then first used to predict the equilibrium sizes of the nanocrystals under different strain conditions using the GWK theorem. We then add an additional layer of geometric complexity – that of O-lattice theory along with coincidence site lattice (CSL) as a special case to visualize the possible dislocation configuration at the interface. These parameters (dislocation line spacing, dislocation vectors) allow the quantification of the effect of the dislocation *i.e.*, release of the elastic interfacial stain energy. It is found that two different growth morphologies of NiO nanocrystals on STO (001) are practically possible - one with positive Wulff point and one with negative Wulff point - depending on whether the interfacial strain is released by dislocation formation or not. This prediction is validated through detailed transmission electron microscopy (TEM) investigations of the NiO/STO samples. We demonstrate that GWK has significant potential as a generalized prediction method for engineered crystal morphology.

**II. Generalised Wulff -Kaishew (GWK) theorem and the predicted equilibrium sizes**

We start with the simple case of a single crystalline island growing on a given surface. The factors that control its growth under various conditions can be primarily simplified in terms of the relative effect of substrate and interfacial strain to surface and/or interface energy[10] shown as

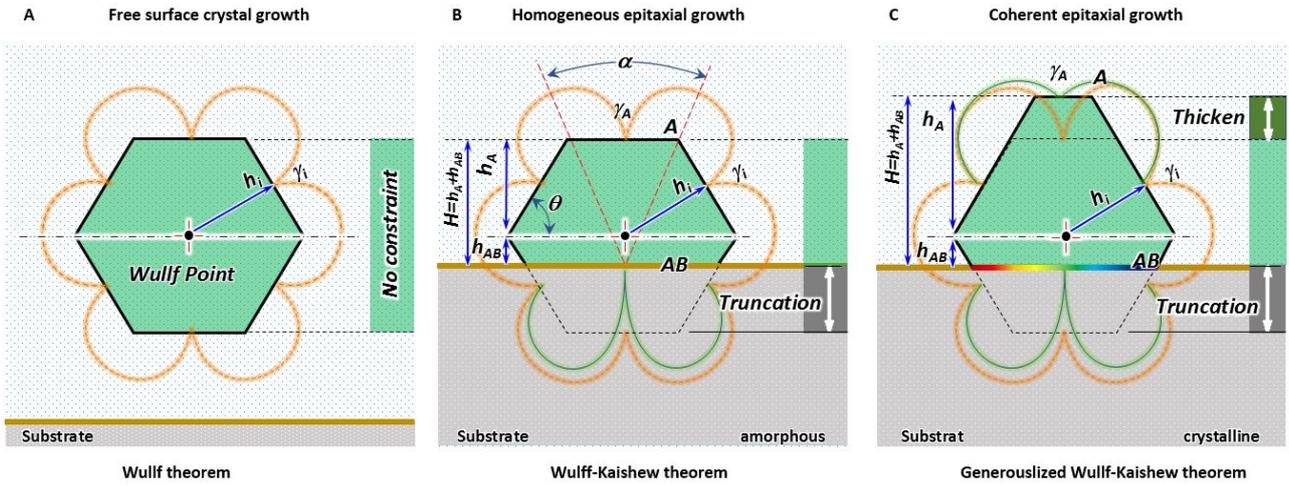

Figure 1. Accordingly this can be split into in three categories, *i.e.*, (a) Free surface growth in

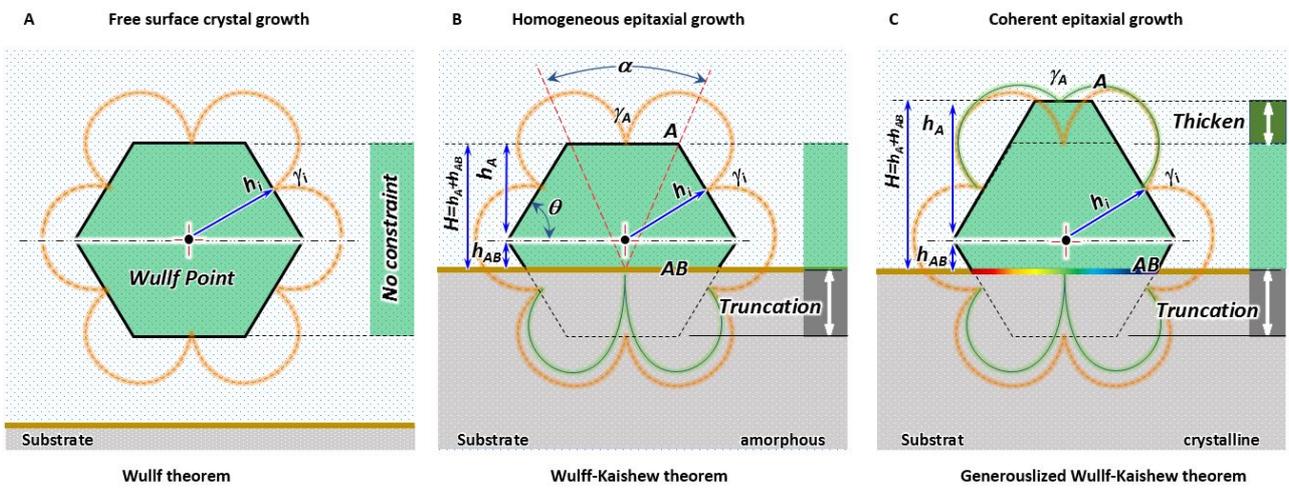

Figure 1A , (b) Homogeneous epitaxial growth without interfacial strain in

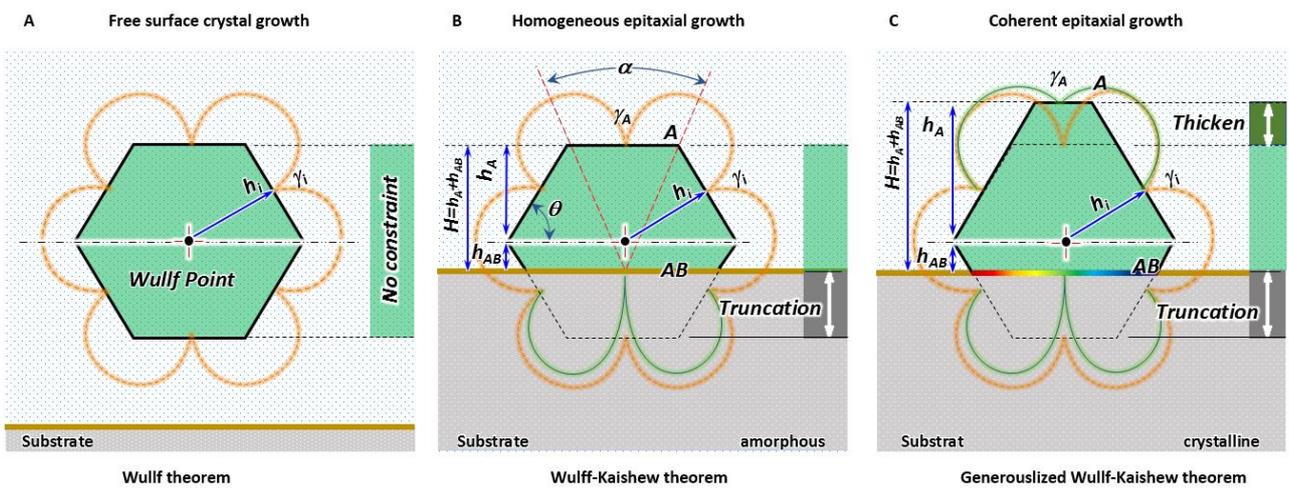

Figure 1B and (c) Coherent epitaxial growth under interfacial strain in

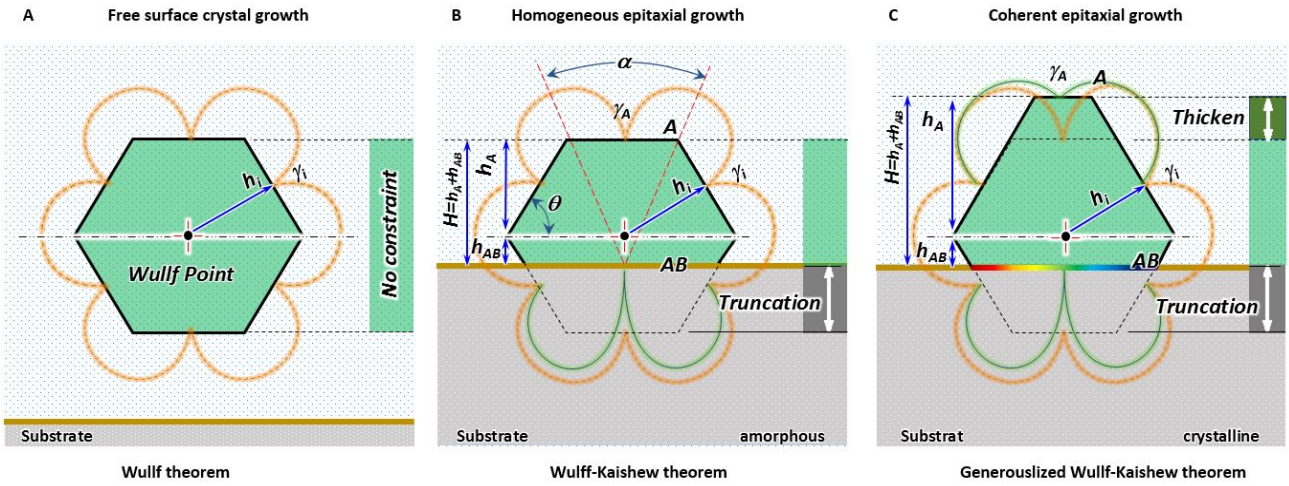

Figure 1C.

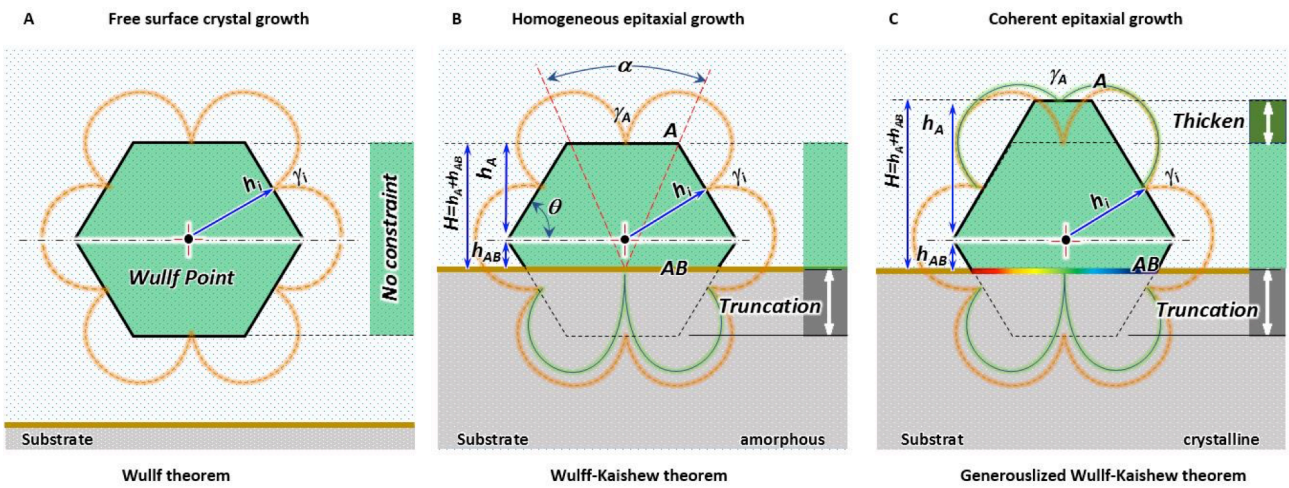

Figure 1 Crystal growth models under variant conditions. A. Free surface growth. B. Homogeneous epitaxial growth and C. Coherent epitaxial growth.

For a free surface crystal growth shown in

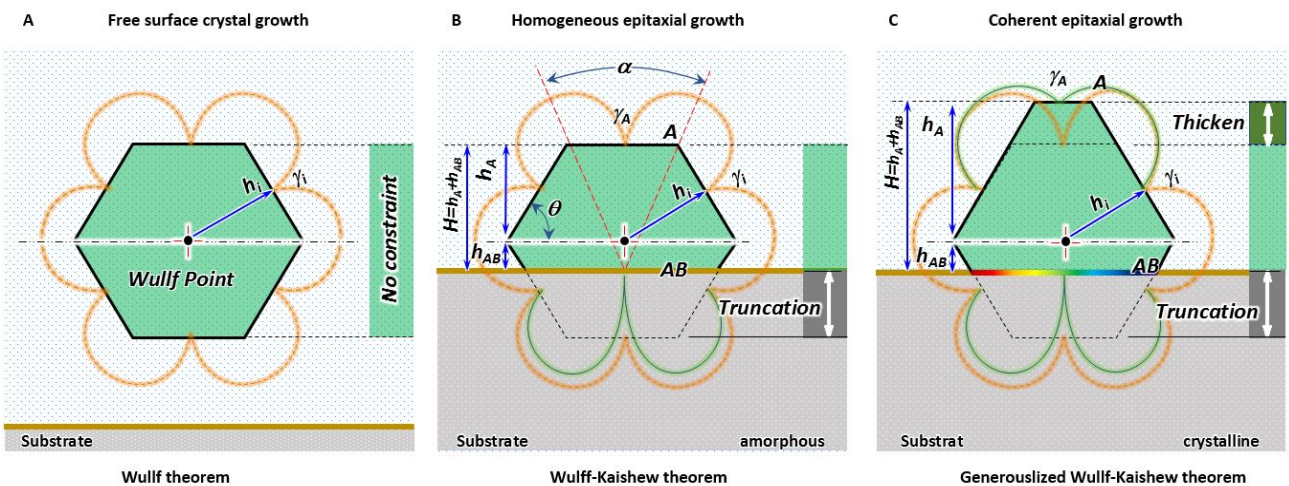

Figure 1A, the growth morphology is only affected by the surface free energy. When reaching the equilibrium shape, the distance of the crystal facets form a point within the crystal (Wulff point) proportional to the corresponding surface energy of each facet, known as Gibbs-Curie-Wulff theorem.[11] For such a crystal, the equilibrium shape is graphically determined using the γ-plot method.[12] The inner envelope of the planes with surface energy $\gamma_i$ perpendicular to directions *i* and proportional to distances $h_i$ measured from the Wulff point (see

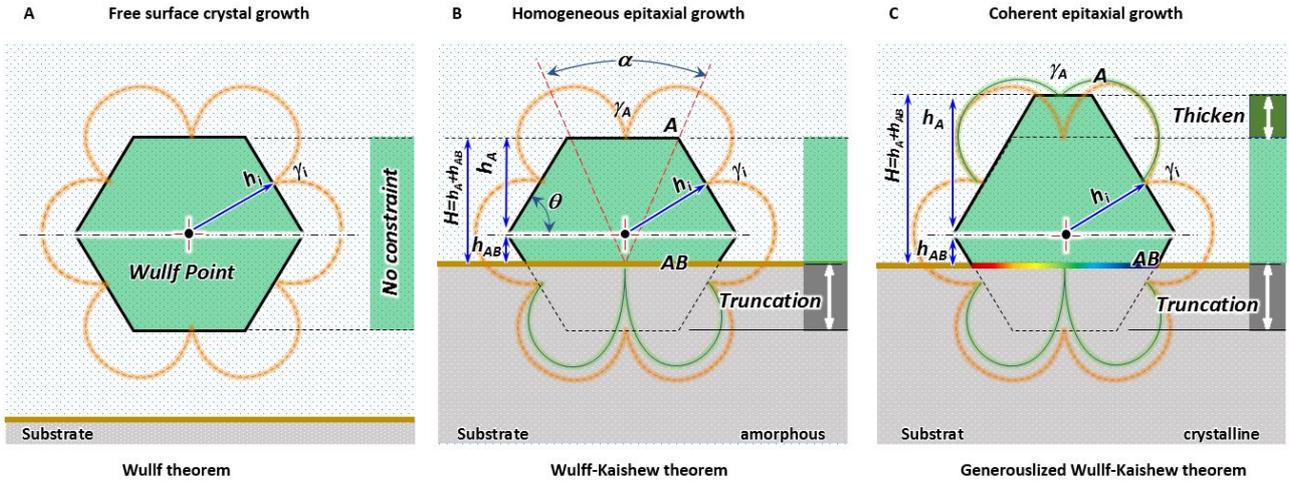

Figure 1A) is governed by:

$$\frac{\gamma_i}{h_i} = constant \qquad 1$$

where $\gamma_i$ is the surface free energy of the $i^{th}$ crystal facet, and $h_i$ is the central distance from Wulff point.

For homogeneous epitaxial growth as shown in

$$\frac{\gamma_{i \neq AB}}{h_{i \neq AB}} = \frac{\gamma_A - \gamma_{adh}}{h_{AB}} = \frac{\gamma_A}{h_A} = \frac{2\gamma_A - \gamma_{adh}}{H} = constant, H = h_A + h_{AB} \qquad 2$$

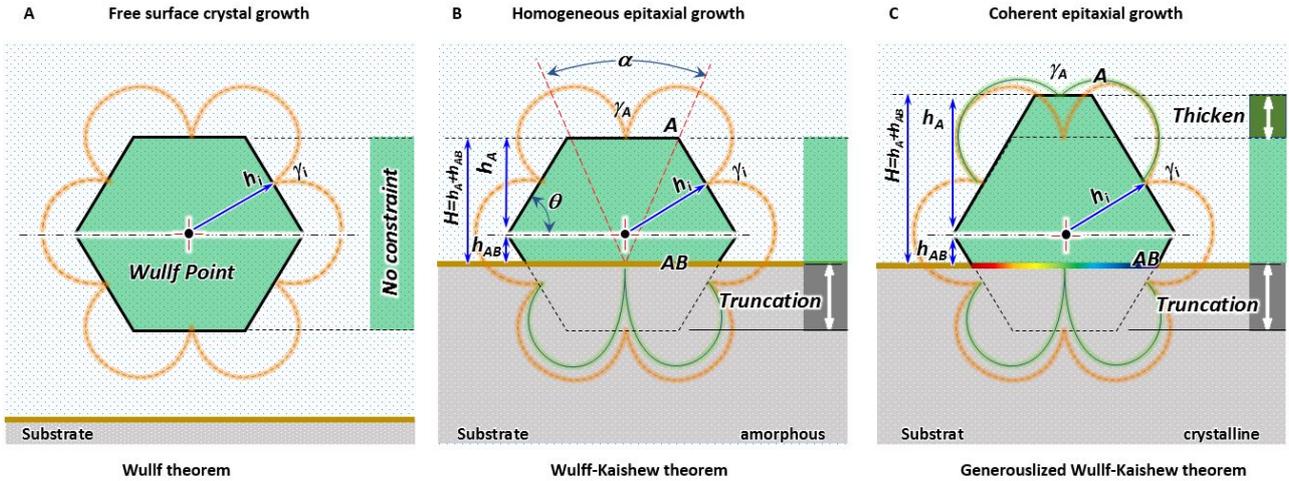

Figure 1B, the substrate supported growth of a nanocrystal without interfacial strain (*i.e.* on top of an amorphous support) can be predicted by a modified Wulff construction, that is Wulff-Kaishew theorem.[13] It has a truncated Wulff construction at the interface. Accordingly, Eq.1 is modified as:

Here '*A*' denotes top surface, '*AB*' stands for interface, $\gamma_A$ is the surface energy of top surface, $\gamma_{adh}$ is the specific adhesion energy of the interface between crystal and the substrate.

$H$ is the crystal height above the interface, $h_A$ is the growth thickness of top surface '*A*', $h_i$ and $h_{AB}$ are the distance of Wulff point to the side facets and bottom surface or interface facet parallel to the top surface, respectively (shown as in

$$\frac{\Delta\mu - \varepsilon_0 m^2 R}{2} = \frac{\gamma_i - \gamma_A \cos\theta_i + \varepsilon_0 m^2 \frac{V}{n_i} \frac{\partial R}{\partial S_i}\big|_{S_{AB}}}{h_i - h_A \cos\theta_i}, \text{(Exposed facets)} \qquad 3$$

$$\frac{\Delta\mu - \varepsilon_0 m^2 R}{2} = \frac{2\gamma_A - \gamma_{adh} + \varepsilon_0 m^2 V \frac{\partial R}{\partial S_{AB}}\big|_{S_i}}{H}, \text{(Substrate truncation)} \qquad 4$$

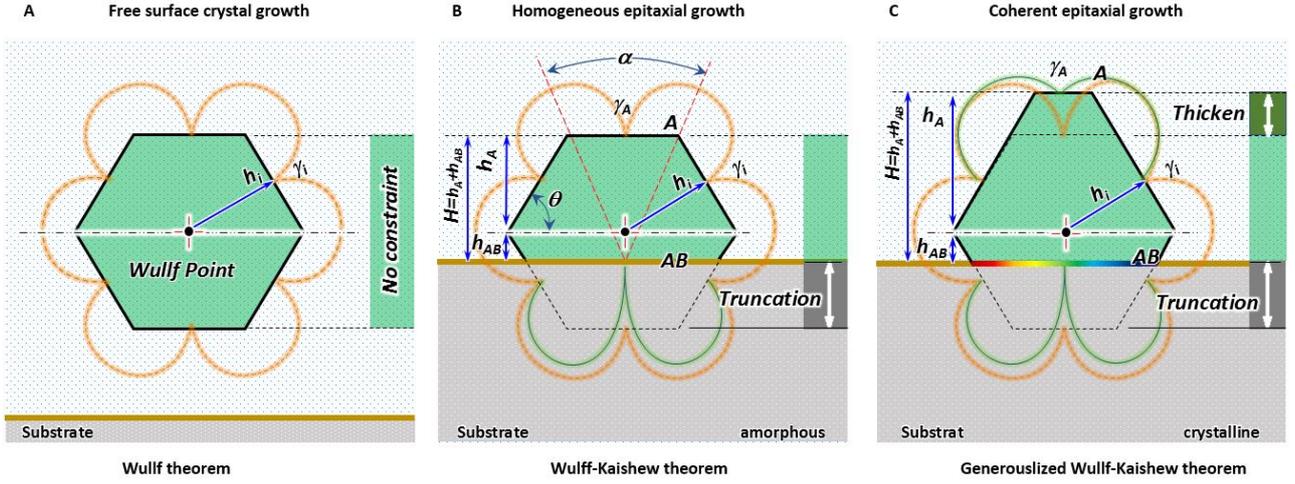

Figure 1B) .

Figure 1C, a nanocrystal grows on a single crystal substrate containing interface mismatch strain. Here the Wulff-Kaishew construction is extended to cover the effect of strain at the interface, named as GWK theorem.[14] In the GWK framework, a symmetric truncated pyramid crystal grown on a single crystalline substrate containing pure elastic interface strain is assumed to reach its equilibrium shape with the height $H_{eq}$ and width $L_{eq}$. In this case, Eq. 2 must be extended as follows:

Where $\Delta\mu$ is the supersaturation per unit column; $\varepsilon_0$ is a combination of the elastic coefficients, equal to $E_A/(1-\nu_A)$, $E_A$ is the Young's module and $\nu_A$ Poisson's ratio; $m$ is lattice misfit between the crystal and the substrate; $m = (a-b)/a$ where $a$ and $b$ is lattice parameters of island crystal and substrate; $R$ the relaxation energy factor in a range of (0, 1), $R = 0$ for completely relaxed crystal and $R = 1$ for no relaxation; $\gamma_A$ and $\gamma_i$ the surface energy of the top surface and the $i^{th}$ facet; $S_i$ is the area of the $i^{th}$ oblique face which included angle with interface is denoted as $\theta_i$ ; $h_i - h_A \cos\theta_i$ a measurement of the top facet $A$ limited by the $i^{th}$ oblique face; $V$ the volume of the symmetric pyramid island crystal.

According to the GGW theorem, the height and the width of the nanocrystal when reaching equilibrium state can be solved from Eqs. 3 and 4 and are defined by the following equations[14]:

$$\begin{cases} H_{eq} = 12\gamma_A \dfrac{\left[tan\left(\frac{\alpha}{2}\right)-\frac{r_0'}{r}(1-2rcot\theta)\right](1-2rcot\theta)}{\varepsilon_0 m^2[3-6rcot\theta+4(rcot\theta)^2]^2}\left|\dfrac{dR}{dr}\right|^{-1} \\ L_{eq} = \dfrac{H_{eq}}{r} \end{cases} \quad 5$$

Where the wetting factor is

$$r_o' = \dfrac{2\gamma_A - \gamma_{adh}}{2\gamma_A} \quad 6$$

The top facet expanding angle parameter measured for Wulff Plot without strain is

$$tan\left(\dfrac{\alpha}{2}\right) = \dfrac{\gamma_1 - \gamma_A cos\theta}{\gamma_A sin\theta} \quad 7$$

The parameters are described as followers:

$\alpha$ - angular extension of the top facet A measured from the Wulff point

$\gamma_{adh}$ - interface adhesion energy

$\theta$ - included angle between the top plane and the side plane of the pyramid.

$r$ - ratio between the height and the width of the pyramid crystal.

$r_o'$ - The wetting factor.

$\gamma_A$, $\gamma_1$ – the surface energy of the top surface and the side surface of the pyramid.

$\varepsilon_0$ – Shear modular equal to $E_A/2(1+\nu_A)$, $E_A$ the Young's module and $\nu_A$ Poisson's ratio.

$m$ – lattice misfit between the crystal (parameter a) and the substrate (parameter b), defined by (b-a)/a.

$R$ – relaxation energy factor in a range of (0, 1). $R = 0$ for complete relaxation and $R = 1$ for no relaxation.

$k(r)$ – The variant of a component of the volume $V = H^3 k(r)/r^2$, $k(r) = 1-2rcot\theta+4(rcot\theta)^2/3$.

For the present system, *i.e.* NiO on STO, we also have to consider gas adsorption that occurs during synthesis as well as misfit dislocations. Where there is gas absorption on side facet inducing extra energy $\gamma_i^{gas}$, interface dislocation array with line spacing d and burger vector b on interface occurring elastic strain partially shrinking to m′ and dislocation energy $\gamma_A^{disl}$, then the Eq. 5 can be extended and shown as below:

$$\begin{cases} H_{eq} = 12\dfrac{\left[tan\frac{\alpha}{2}-r_0'\frac{1-2rcot\theta}{r}\right](\gamma_A+\gamma_A^{gas})(1-2rcot\theta)}{\left(\varepsilon_0 m'^2 + \gamma_A^{disl}\right)K^2(r)}\left|\dfrac{dR}{dr}\right|^{-1} \\ L_{eq} = \dfrac{H_{eq}}{r} \end{cases} \quad 8$$

where

$$\text{Remaining interfacial surface } m' = m - N|\mathbf{b}|/\sqrt{S_{AB}} = m - |b|/d, \quad 9$$

Where $N$ is the number of interface dislocation lines and d is the equilibrium dislocation d-spacing

$$r_0' = \dfrac{2\gamma_A - \gamma_{adh} + \gamma_A^{gas}}{2(\gamma_A + \gamma_A^{gas})} \quad 10$$

$$tan\dfrac{\alpha}{2} = \dfrac{(\gamma_i + \gamma_i^{gas}) - (\gamma_A + \gamma_A^{gas})cos\theta_i}{(\gamma_A + \gamma_A^{gas})sin\theta_i} \quad 11$$

$$K(r) = 3 - 6rcot\theta_i + 4r^2 cot^2\theta_i \quad 12$$

When there is no gas adsorption on side facet but only interface dislocation array with line spacing *d* on interface occurring dislocation energy $\gamma_A^{disl}$, then the Eq. 8 can be simplified below:

$$\begin{cases} H_{eq} = 12\dfrac{\left[tan\frac{\alpha}{2}-r_0'\frac{1-2rcot\theta}{r}\right]\gamma_A(1-2rcot\theta)}{\left(\varepsilon_0 m'^2 + \gamma_A^{disl}\right)K^2(r)}\left|\dfrac{dR}{dr}\right|^{-1} \\ L_{eq} = \dfrac{H_{eq}}{r} \end{cases} \quad 13$$

**III. Prediction of morphology of NiO crystal epitaxial grown on STO (001) substrate**

From Eq. 13, we find that the morphology of NiO nanocrystals epitaxially grown on STO (001) substrate is mainly affected by three important parameters. The first parameter is the ratio between the released interfacial strain and the full coherent elastic strain. It is defined by $C1 = (m-m')/m$. $C1$ is 0 for fully coherent and is 1 when elastic strain is fully accommodated by interfacial dislocation. The second parameter affecting the NiO morphology is the surface energy ratio between side facet and top facet, which is expressed as $C2 = \gamma_I/\gamma_A$. This is important, as often the exposed surfaces during epitaxial growth may be altered by the growth environment (thermo gradient, gas absorption, etc.). The third parameter, the ratio between growth height (H) and width (L), $r = H/L$, is used for defining growth morphology.

Figure 2 illustrates the systematic evolution of growth morphology of NiO in terms of alterable variants $r$, $C1$ and $C2$. The coordinates (x, y) of the points at the upper blue curve ($P_{UB}$), the lower blue one ($P_{LB}$), the upper red ($P_{UR}$) and the lower red ($P_{LR}$) are geometrically illustrated in Figure 2 D. All the curves are generated by using the above equations.

With the simplest case where only elastic strain occurs at the interface, it is found that the Wulff shape takes negative Wulff point (white solid circles underneath the interface between the crystal and the interface, see Figure 2 A). If the ratio r changes, the truncated pyramid shape will slowly turn into a triangular shape at a certain ratio. It indicates that width of the top facet will slowly increase followed by a rapid vanishing from $r = 0.6$ and eventually the growth morphology of the nanocrystals with equilibrium Wulff shape collapses into triangular prism.

When the interfacial elastic strain starts to be accommodated by induced dislocation defects, the Wulff point starts to approach to positive position with increasing ratio between residual elastic strain and full elastic strain ($C1$ varies from 0 to 1) as shown in Figure 2B. This shows equilibrium shape evolution under the interface strain and dislocation effect. It is interesting to find that pyramid shape is always truncated, however, the size of the crystal does not continue to grow infinitely; indeed, the interfacial strain plays a role in limiting the growth of NiO.

Since surface energy for both top and side facets may vary during growing under certain circumstances (especially for a vapour-solid or hydrothermal growth), the effect of the surface energy ratio between side facet and top facet ($C2$) is also explored and shown in Figure 2 C (for the case of a dominant elastic strain interface). It finds that when the ratio $C2$ increases, the NiO crystal thickens quickly and becomes narrower. Thus, the comparison of the profiles indicates that:

i) Under equilibrium conditions, the crystal size can continue to coarsen but will reach a limit when r approaches to around 0.6. Figure 2A is the case for $r = 0.52$ corresponding to a height ~80 nm for light blue polygonal NiO.

ii) When there exists a certain interfacial misfit energy ($0 < C1 < 1$), the self-similar feature of the morphology is lost and the Wulff point O changes from negative to positive position along the normal of the interface. When the ratio r linearly increases under a certain interface strain energy, the truncation occurs at the beginning and quickly reaches to the maximum lateral size followed by shrinking until vanishing with further increase of ratio r.

iii) The ratio of surface energy between side facet and top facet, $C2$, modifies crystal shape but the effect to the Wulff shape is much smaller than that of interface strain ratio between elastic strain and plastic strain, $C1$. Obviously, $C1$ quickly lifts the Wulff point from negative to positive position while $C2$ just slightly pushes the Wulff point up but does not change the sign the Wulff point.

**IV Experimental results and discussion: TEM observation of NiO morphology and the interface dislocation**

To test the theory developed in Section 2, we investigated NiO nanostructures, 10 ~ 100 nm wide and up to 30 nm high epitaxially grown on (001) STO substrates. The synthesis details can be found from elsewhere.[15] Here we present two cases of NiO crystals that grow on STO (001) surface but with opposing situations with respect to the Wulff point. Case I has a negative Wulff point and Case II has a positive Wulff point. If we disregard surface gas adsorption, the only difference of these two cases lies in the magnitude of the interfacial strain. Case I has no dislocation strain energy at the interface and hence dominated by elastic mismatch strain energy. In contrast, for Case II all the interfacial strain has been relaxed by misfit dislocations thus there is a domination of interfacial dislocation energy. The crystallographic structure and the lattice parameter of NiO and STO for calculation can be found in

Table 1.

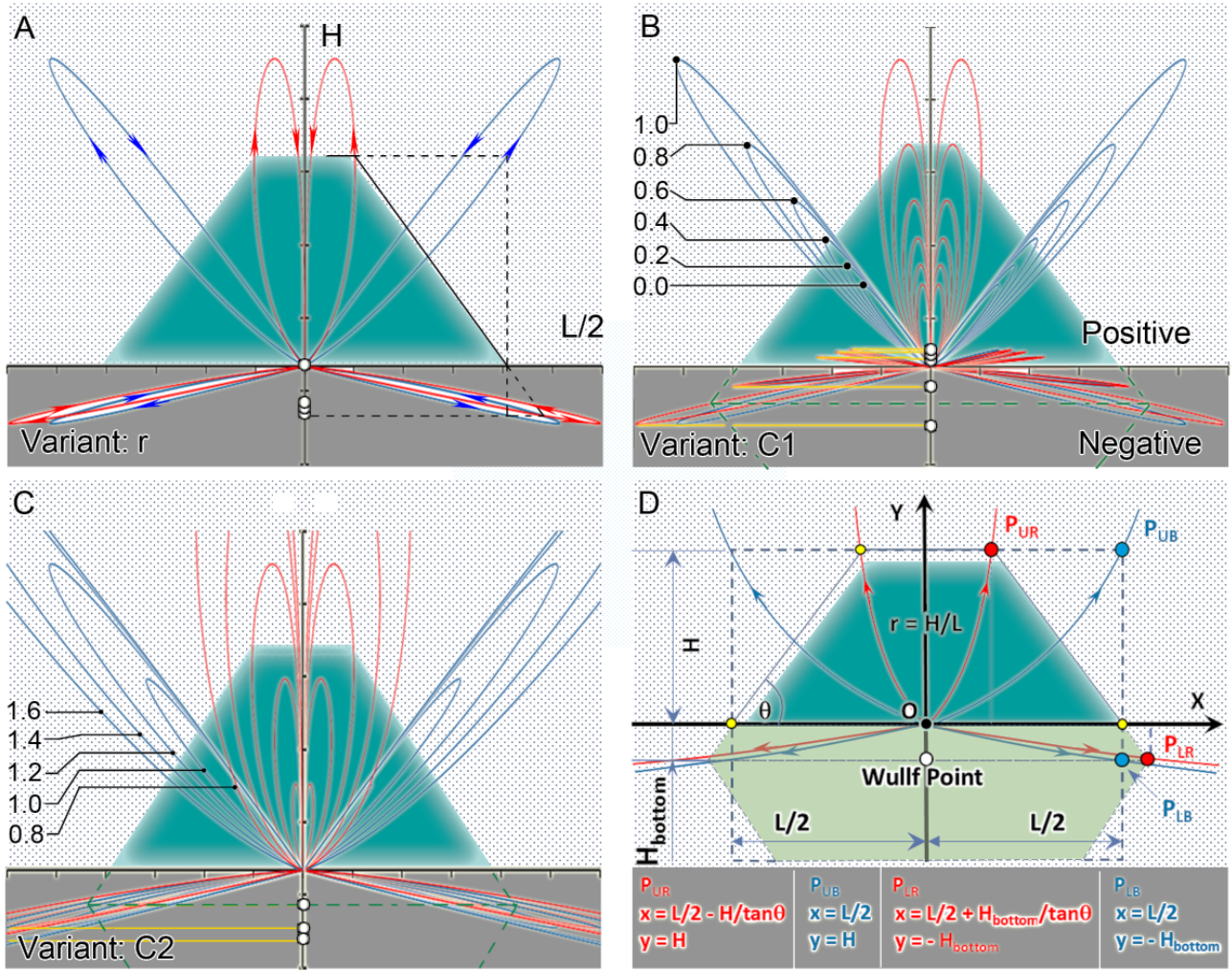

Figure 2 Equilibrium shape of truncated pyramid NiO nanocrystal growing on STO (001) substrate calculated for continuous curves and the edges trajectories. A. Equilibrium shape changes from truncated pyramid to full pyramid when both height-to-length ratio and the length dimension linearly increasing with existence of interface strain. B. Equilibrium shape changes within truncated pyramid when *C2* as variant. Wulff point changes from negative to positive with increasing dislocation lines occurring at interface. C. Equilibrium shape changes within truncated pyramid when *C1* as variant. The coordinates scale is in nm.

**Table 1 Initial OR and lattice parameters of NiO and STO**

| Phase | NiO (*p*) | STO (*m*) |
|---|---|---|
| Crystal Structure | FCC | PC |

| Space Group | $Fm\bar{3}m$ (225) | $Pm\bar{3}m$ (221) |
|---|---|---|
| Strukturbericht symbol | B1 | E21 |
| Lattice Parameter | $a_p$ = 0.4168 | $a_m$ = 0.3905 |
| Parallel vectors | $[010]_p$ | $[010]_m$ |
| Parallel planes | $(100)_p$ | $(100)_m$ |

**Case I: NiO with negative Wulff point**

The morphology of NiO nanocrystal grown on STO (001) surface is as shown in Figure 3A. The NiO nanocrystal and STO are on cube-on-cube orientation relationship, as shown in Fast Fourier Transformation (FFT) of Figure 3A (Figure 3B), identical to the reported rock-salt type nickel oxide.[16] The overall growth morphology of NiO nanocrystal is enclosed by {001} facets at the interface, {111}, {113} and trace amount of {110}.

The surface energy of NiO is relatively low for planes {001}, {111} and {113}, therefore the measured growth size of these low-surface-energy and low-growth-speed planes are very similar to each other. On the other hand, the interfacial adhesion energy of (001) NiO and (001) STO is very high and consequently NiO nanocrystal are growing with a negative depth along the normal of (001) interface, which leads to a negative Wulff point. The distance between the facet and the Wulff Point of NiO is ~17 nm for {001}, 23 nm for {011}, 20 nm for {111} and 19.8 nm for {113} respectively. The growth height is 17.6 ± 0.5 nm and the width is 46.3 ± 0.5 nm. The ratio between height and width is 0.38. The overall 3D morphology of NiO takes a near spherical shape but deeply imbedded into STO substrate with only about one-third exposed towards to the outer side as shown in Figure 3C&D.

Critically no misfit dislocation lines were observed at the interface of NiO/STO in the Figure 3A, which signifies this growth is mainly controlled by elastic mismatch strain. That is the residual mismatch $m' = m$ and dislocation energy equals to zero. Recall that in Figure 2C, we have shown the effect of the variant $C2$ on the growth morphology. We find that the predicted morphology is close to our observations for $C2$ values ~ 1.2. Physically, this means the practical surface energy of (111) plane is 1.2 times of that of the plane (001) as a result of the growth of the vapor environment. The height of NiO at equilibrium state could be compared to those predicted by the GWK theorem shown in Eq. 5 and Table 2. We obtain an equilibrium height of 15.74 nm (details of the calculation shown in Appendix S4.1) in close agreement with the experiment.

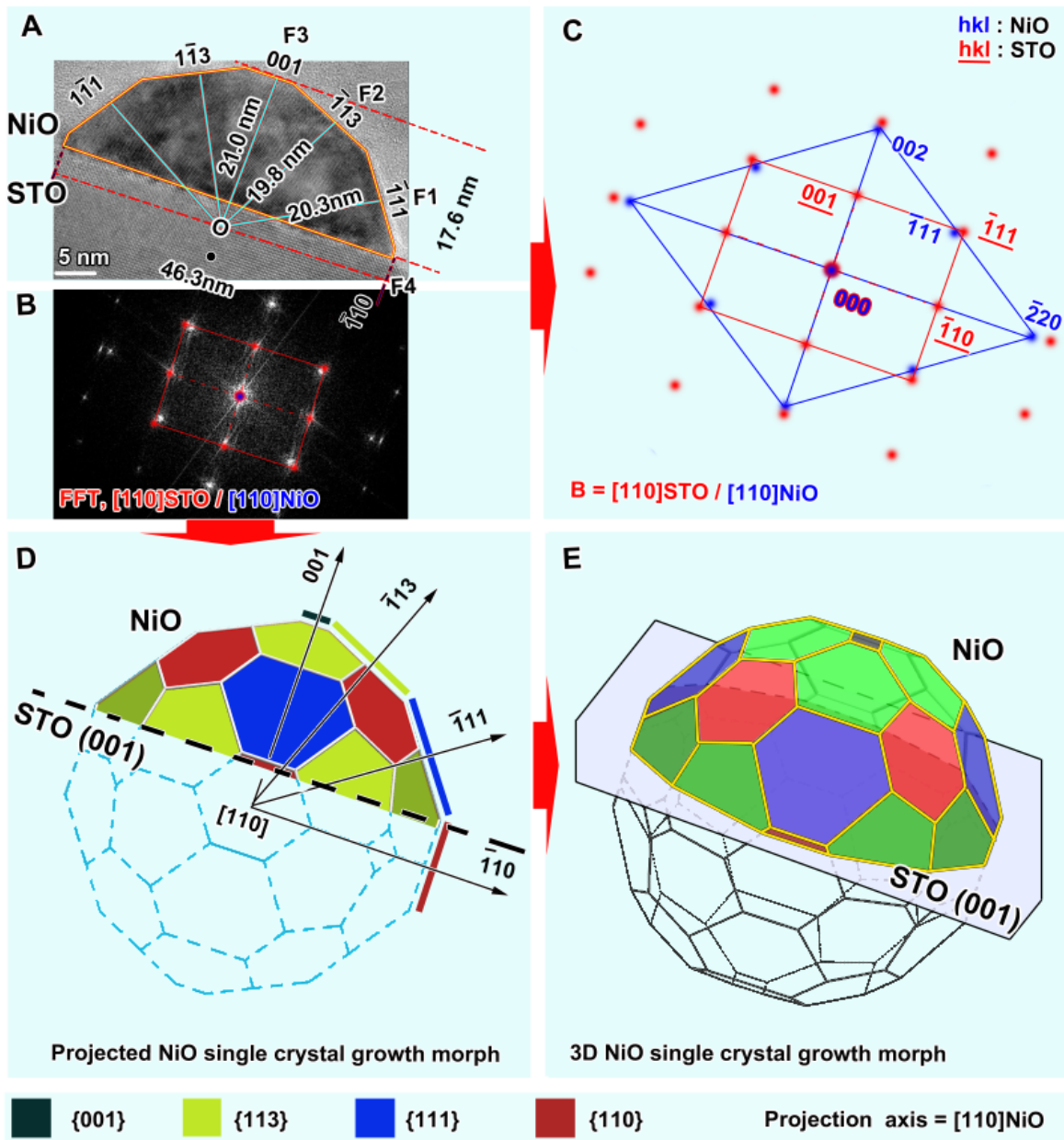

Figure 3 The growth morphology of single crystal stoichiometric NiO on STO (001) surface. A. High resolution TEM image viewed at [110] STO and [110] NiO orientation which is the edge-on direction of the STO (001) substrate surface. Please note that each exposed facet has been indexed with both plane and growth depth. The FFT image (Panel B) of the grown NiO crystal in Panel A and could be indexed as [110] NiO zone axis diffraction pattern (Panel C). It indicates the lattice parameter of NiO is close to 0.417 nm, identical to so-called rock-salt structure of NiO. D. A polymorph of single crystal NiO projected along [110] direction with negative Wulff point enclosed by the exposed facets {001}, {011}, {111} and {113}, coloured in dark blue, light green, light blue and dark red respectively. E. A 3D model of morphology of NiO crystal shows a typical ball-crown shape with a near smooth

**Case II: NiO with positive Wulff point**

The second case we investigate is of those nanocrystals that show a positive Wulff point. An example is shown in Figure 4A. This happens when the exposed facets of the above planes change during the growth of NiO crystal such that as it approaches to near equilibrium status, and only {001} and {111} facets remain. This is corresponding to square shape of NiO crystals when viewed at [001] direction as shown in the ref. [15b]. Note that distinct interfacial dislocation lines are observed for such well-developed NiO crystals with positive Wulff-point.

As shown in Figure 4A, the nanocrystal takes the shape of a typical truncated quadrangular pyramid enveloped by a dominant {111} facet with trace amount of {001}, *i.e.* the{113} facets have vanished. The Wulff point is above the interface level. The measured height and width of NiO nanocrystal is around 34.0 nm and 54.7 nm separately. The ratio between the height and the width is 0.62.

The overall 3D morphology of NiO takes a near spherical shape but deeply imbedded into STO substrate with only about one-third exposed towards to the outer side as shown in Figure 3C&D.

Such nanocrystals with truncated pyramidal shape are only seen when the interface strain is fully relaxed by misfit dislocation array formation. A detailed crystallographic calculation of interfacial dislocation structure is presented in Appendix S3. It indicates that the dislocation line at the STO (001) / NiO (001) interface is [110]/2 with spacing $d_0$ = 4.38 nm at its equilibrant status. The calculated value is very close to he observed d-spacing shown in Figure S1A. Hence interfacial mismatch strain has been fully accommodated at the interface between NiO and STO by an array of dislocation lines with d-spacing of around 4 nm. It indicates a full release of mismatch strain by an equilibrium dislocation lines at the interface. This dislocation energy of a dislocation array with d-spacing $d$ = 4.38 nm has been solved as $\gamma_A^{disl}$ = 0.529 J/m2 (See Appendix S4.2). This now allows us to compute the equilibrium height of the NiO islands for case II (*i.e.* nanocrystals with positive Wulff point) for the given growth ratio. Full details of calculations are shown in the Appendix S4.2.

Table 2 is a list of all the parameters and the computation results for the both cases and the comparison with experimental data. We find that for each case the calculated height for NiO reaching equilibrium state is very close to the actual growth height in the experiment. This indicates that the GWK construction and methodology is effective in predicting as well as rationalizing the shape and size features of epitaxial NiO nanocrystals.[17]

**Table 2 A summary of surface energy measurement and calculation for case I and II nanocrystals on STO (001) substrate via GWK construction**

| Item | Morphology parameters | | | | Energy parameters | | | | Intermediate variants | | | | | | Theoretic result | |
|---|---|---|---|---|---|---|---|---|---|---|---|---|---|---|---|---|
| | $H$ | $L$ | $h_{top}$ | $h_{bottom}$ | $\gamma_A$ | $\gamma_1$ | $\gamma_{adh}$ | $\gamma^*$ | $m'$ | $r_{exp.}$ | $r_{min}$ | $r_{max}$ | $r_o'$ | $dR/dr$ | $H_{eq.}(r_{exp.})$ | Dev. |
| | nm | nm | nm | nm | J/m² | J/m² | J/m² | J/m² | * | * | * | * | * | * | nm | % |
| Case I | 17.6 | 46.3 | 21.0 | -3.4 | 1.02 | 1.22 | 1.24 | -0.16 | 0.063 | 0.380 | 0.297 | 0.707 | 0.392 | -0.271 | 15.7 | 10.8 |
| Case II | 31.5 | 54.7 | 34.0 | 2.6 | 1.02 | 1.22 | 1.24 | 0.28 | 0 | 0.621 | 0.297 | 0.707 | 0.392 | -0.050 | 30.0 | 4.76 |

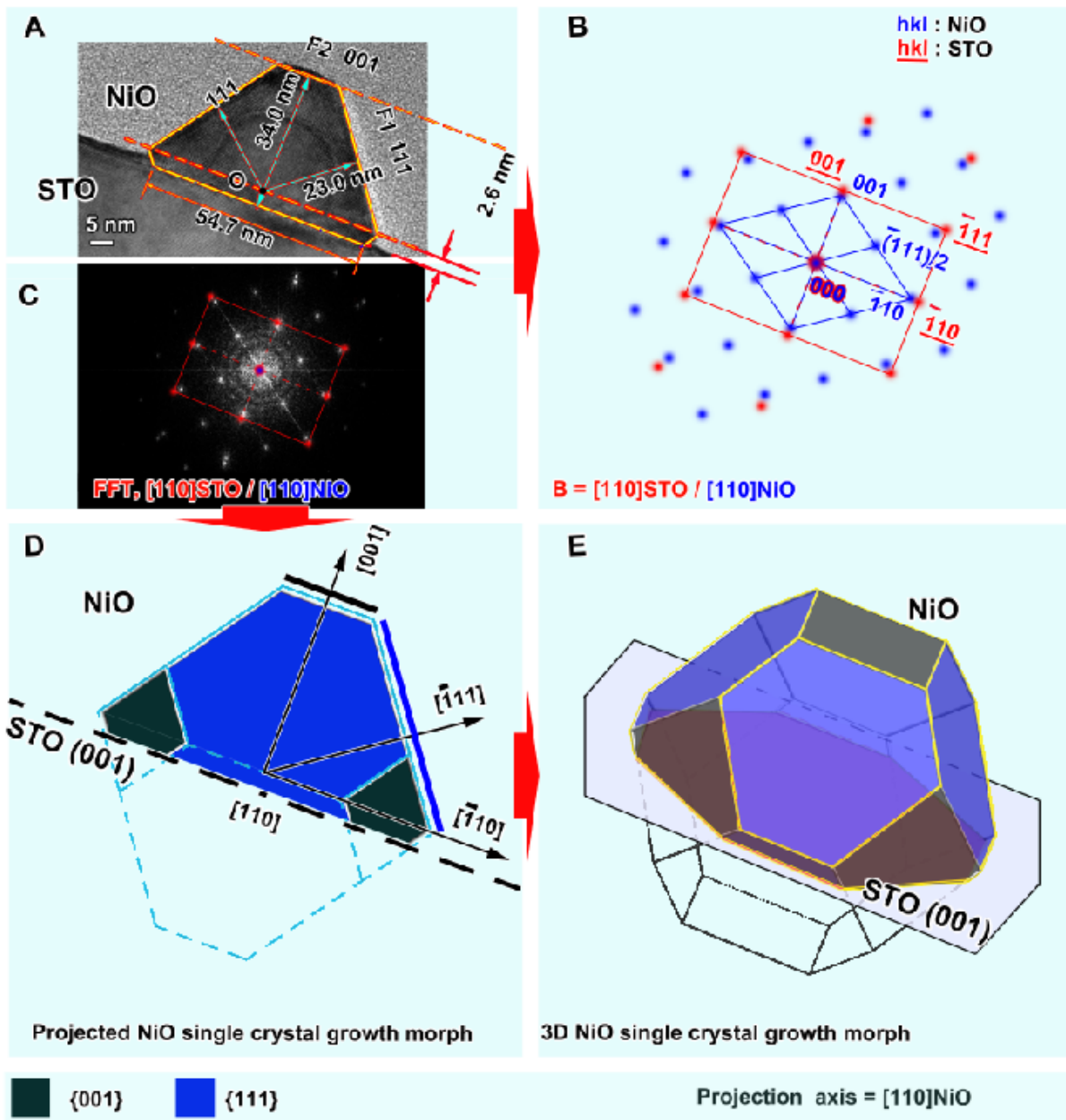

Figure 4 The well-developed growth morphology of single crystal NiO on STO (001) surface. A. High resolution TEM image viewed at [110] STO and [110] NiO orientation which is the edge-on direction of the STO (001) substrate surface. Please note that each exposed facet has been indexed with both plane and growth depth. The viewed direction is revealed by the corresponding FFT image (Panel C). B. The indexing result of the FFT in Panel C. The FFT image of the grown NiO crystal in Panel A and could be indexed as [110]NiO zone axis diffraction pattern. D. A polymorph of single crystal NiO projected along [110] direction with negative Wulff point enclosed by the exposed facets {001} and {111} coloured in dark blue and light blue respectively. E. a 3D model of morphology of NiO crystal shows a typical truncated quadrangular pyramid shape with a near flat surface.

## V Conclusion

In summary a study of the factors that control the shape of epitaxial NiO individual nanocrystals on STO (001) substrate has been presented under the framework of a generalised Wulf-Kaishew theorem. Two scenarios were predicted and experimentally confirmed. In scenario I the NiO nanocrystal has a negative Wulff point as a result of a fully coherent epitaxially strained interface with the underlying substrate. Here the NiO takes a smooth ball-crown morphology with {001},

{011}, {111} and {113} exposed facets. In scenario II, the NiO nanocrystal has a positive Wulff point. Here the interfacial elastic strain has been relaxed by dislocation arrays with d-spacing about 4.38 nm. A constraint displacive shift complete pattern is found, defined by two in-plane vectors am [011]/2 and $a_m[01\bar{1}]/2$, which are the shortest Burgers vectors of the dislocation lines lying on the interface. Such nanocrystals develop a truncated quadrangular pyramid morphology as their equilibrium shape. There is a very good agreement between the shape and dimensions predicted by the GWK framework and the experimental analysis performed by high resolution TEM. This validation indicates that this method of analysis can be extended to a large variety of functionally important oxide nanocrystal growth cases.

We end the manuscript with a brief discussion on the importance of vacuum environments. One of our key conclusions is that one must also recognize that the general theorems used to predict crystal morphology work best when the fabrication occurs in ultra-high vacuum environments. This is not necessarily the case for oxide nanocrystals where often fabrication is done under a certain partial oxygen pressure. This partial gas pressure takes a very important role when accounting for the free surface energy. For example, following the results of Oliver et al,[18] we expect that surface termination would have a definite influence on the {111} and {113} surfaces but not on {100} and {110} ones. This results in an differing growth speed for the bounding surfaces resulting asymmetric crystal shape. Fortunately, we did observe such a morphology. Figure S6 in Appendix S5 is an example for stoichiometric NiO grown at the same condition of that shown in Figure 3A. In this image, both {111} and {113} have asymmetric growth which results in a highly irregular crystal shape. Having said that,

(i) the calculated results in Table 2 (which only considers interface strain and dislocations) is already very close to the experimental observation with a deviation for each case is only ~10% and,

(ii) such nanocrystals with asymmetric growth signatures were relatively rare amongst the several hundred odd nanocrystals we investigated. It suggests that the interface strain plays a more dominant role in determining the growth morphology, at least for this system when compared to oxygen partial pressure.

From a numerical perspective when NiO crystal grows in partial oxygen pressure, gas adsorption contributes to the decrease $\Delta\gamma^{gas}$ of the surface energy in Eq. 10 and 11. A few attempts have already been made[19,4b] but the practical difficulties of measurement of gas adsorption is a significant and challenging problem. The incorporation of the effect of oxygen adsorption on the equilibrium shape analysis of an epitaxial nanocrystals in the GWK framework forms part of our future work.

## ASSOCIATED CONTENT

**Supporting Information**

The Supporting Information is available free of charge on the ACS Publications website. (file type, PDF)

## AUTHOR INFORMATION

**Corresponding Authors**

[*] E-mails: hongwei.liu@sydney.edu.au;
          nagarajan@unsw.edu.au

**Present Addresses**

[†] Department of Biochemistry & Molecular Biology, Monash University, Melbourne, VIC 3800, Australia

**Author Contributions**

[‡] These authors contributed equally

## ACKNOWLEDGEMENT

The authors acknowledge the facilities and the scientific and technical assistance of the Australian Microscopy & Microanalysis Research Facility at the Sydney Microscopy & Microanalysis (SMM), The University of Sydney. X. Cheng and V. Nagarajan acknowledge financial support from the ARC Discovery Project and Dr. Kashinath Bogle and Jivika Sullaphen for the nanocrystal synthesis.